\pdfoutput=1
\documentclass[11pt,a4paper]{article}
\usepackage{amsmath,amssymb,url,hyperref,graphicx,cite, color, listings}

\usepackage{tikz}
\usetikzlibrary{shapes, arrows}

\begin{document}
\rightline{DESY 19-148}

\begin{center}
{\large \bf 
\texttt{SimpleBounce} : a simple package for the false vacuum decay
}

\vskip 1.2cm
Ryosuke Sato
\vskip 0.4cm

{\it Deutsches Elektronen-Synchrotron (DESY), Notkestra\ss e 85, D-22607 Hamburg, Germany}

\vskip 1.5cm

\abstract{
We present \texttt{SimpleBounce}, a C++ package for finding the bounce solution for the false vacuum decay.
This package is based on a flow equation which is proposed by the author \cite{Sato:2019axv}
and solves Coleman--Glaser--Martin's reduced problem \cite{Coleman:1977th}:
the minimization problem of the kinetic energy while fixing the potential energy.
The bounce configuration is obtained by a scale transformation of the solution of this problem.
For models with 1--8 scalar field(s),
the bounce action can be calculated with ${\cal O}(0.1)$ \% accuracy in ${\cal O}(0.1)$ s.
This package is available at \url{https://github.com/rsato64/SimpleBounce}.
}
\end{center}

\begin{description}
\item [Program title:]~\\ SimpleBounce
\item [Licensing provisions:]~\\ GPLv3
\item [Programming language:]~\\ C++
\item [Nature of problem:]~\\
The decay rate of the false vacuum can be evaluated by using Euclidean path integral formalism,
and the bounce solution is a saddle point of this path integral.
The bounce solution and its Euclidean action are numerically evaluated.
\item [Solution method:]~\\
The bounce solution is obtained as a fixed point of gradient flow equation.
\item [Additional comments including Restrictions and Unusual features:]~\\ The computation is fast and stable against a choice of initial configuration.
\item [References:]~\\
R.~Sato, ``{Simple Gradient Flow Equation for the Bounce Solution},''
  \href{http://dx.doi.org/10.1103/PhysRevD.101.016012}{{\em Phys. Rev. D}
  {\bfseries 101} no.~1, (2020) 016012},
  \href{http://arxiv.org/abs/1907.02417}{{\ttfamily arXiv:1907.02417
  [hep-ph]}}.
\end{description}

\section{Introduction}
The decay of the false vacua is an important phenomenon in particle physics and cosmology.
The lifetime of the false vacua can be calculated by the Euclidean path integral \cite{Coleman:1977py}.
(For an earlier work, see Ref.~\cite{Lee:1974ma, Frampton:1976kf}.)
In this formalism, the bounce solution which is a saddle point of the action gives the dominant contribution to the decay width.
There exist several numerical packages to calculate the bounce solution;
\texttt{CosmoTransitions} \cite{Profumo:2010kp, Wainwright:2011kj}\footnote{
\url{https://clwainwright.github.io/CosmoTransitions/}},
\texttt{AnyBubble} \cite{Masoumi:2016wot}\footnote{
\url{http://cosmos.phy.tufts.edu/AnyBubble/}},
and \texttt{BubbleProfiler} \cite{Akula:2016gpl, Athron:2019nbd}\footnote{
\url{https://github.com/bubbleprofiler/bubbleprofiler}}.
For other numerical algorithms, see Refs.~\cite{Claudson:1983et, Kusenko:1995jv, Kusenko:1996jn, Moreno:1998bq, Cline:1998rc, John:1998ip, Cline:1999wi, Konstandin:2006nd, Park:2010rh, Guada:2018jek, Jinno:2018dek, Espinosa:2018szu, Piscopo:2019txs}.

In this paper, we introduce a new numerical package to calculate the bounce solution.
Our package utilizes a gradient flow equation. This framework is recently proposed by Chigusa, Moroi, and Shoji \cite{Chigusa:2019wxb}.
Our package solves a flow equation which is proposed by the author \cite{Sato:2019axv}.
As we will see in the next section, our flow equation minimizes the kinetic energy of configuration while fixing the potential energy.
Thanks to Coleman--Glaser--Martin's discussion \cite{Coleman:1977th},
the bounce solution is obtained by a scale transformation of the fixed point of the flow equation.

\section{Formulation}
Here we briefly summarize the gradient flow equation which is proposed in Ref.~\cite{Sato:2019axv}.

We take the Euclidean action with $n_\phi$ scalar fields which have the canonical kinetic term and a generic potential term:
\begin{align}
{\cal S}_E[\phi] = {\cal T}[\phi] + {\cal V}[\phi], \qquad
{\cal T}[\phi] = \sum_{i=1}^{n_\phi} \int d^d x \frac{1}{2} (\nabla \phi_i)^2, \qquad
{\cal V}[\phi] = \int d^d x [V(\phi)-V(\phi_{FV})].
\end{align}
Here $d$ is the dimension of the Euclidean space, and we assume $d$ is larger than 2.
The scalar potential $V$ satisfies $\partial V/\partial\phi_i |_{\phi = \phi_{FV}} = 0$,
all of the eigenvalues of the Hessian of $V$ at $\phi_i = \phi_{FV,i}$ are non-negative,
and $V(\phi) - V(\phi_{FV})$ is somewhere negative.
The bounce solution is a configuration which satisfies the following equation of motion and the boundary condition at infinity:
\begin{align}
-\nabla^2\phi_i + \frac{\partial V}{\partial \phi_i} = 0, \qquad 
\lim_{|x|\to\infty} \phi_i(x) = \phi_{FV,i}.
\end{align}
The bounce solution has $O(d)$ symmetry \cite{Coleman:1977th, lopes1996radial, byeon2009symmetry, Blum:2016ipp} in the space.
Thus, the equation of motion can be simplified as
\begin{align}
-\frac{d^2\phi_i}{dr^2} - \frac{d-1}{r} \frac{d\phi_i}{dr} + \frac{\partial V}{\partial \phi_i} = 0.  \label{eq:EOM}
\end{align}

Here we solve the minimization problem of the kinetic energy ${\cal T}[\phi]$ while fixing the negative potential energy ${\cal V}[\phi]<0$
rather than solving the above equation of motion directly.
This is the reduced problem of the bounce solution which is proposed in Ref.~\cite{Coleman:1977th}.
The solution of this problem is a scale-transformed of the bounce solution.
In Ref.~\cite{Sato:2019axv}, a gradient flow equation is proposed to solve this problem.
We introduce functions $\varphi_i(r,\tau)$ and the flow of $\varphi$ is described as
\begin{align}
\frac{\partial}{\partial \tau} \varphi_i(r,\tau) &= \nabla^2 \varphi_i - \lambda[\varphi] \frac{\partial V(\varphi)}{\partial \varphi_i} \label{eq:gradient flow equation}, \\
\lambda[\varphi] &= \frac{\displaystyle\sum_{i=1}^{n_\phi} \int_0^\infty dr r^{d-1} \frac{\partial V(\varphi)}{\partial \varphi_i} \nabla^2 \varphi_i}
{\displaystyle\sum_{i=1}^{n_\phi} \int_0^\infty dr r^{d-1} \left(\frac{\partial V(\varphi)}{\partial \varphi_i}\right)^2}.
\label{eq:lambda}
\end{align}
Note that $\lambda[\varphi]$ depends on $\tau$ but \textit{not} on $r$.
Here $\tau$ is ``time'' for the flow of $\varphi$ and $\nabla^2 \varphi_i = \partial_r^2 \varphi_i + (d-1) (\partial_r \varphi)/r$.
We take the initial $\varphi(r,0)$ such that
\begin{align}
{\cal V}[\varphi] |_{\tau=0} < 0. \label{eq:initial V}
\end{align}
We can easily check
\begin{align}
\frac{d}{d\tau}{\cal V}[\varphi] = 0, \qquad
\frac{d}{d\tau}{\cal T}[\varphi] \leq 0. \label{eq:evolution V and T}
\end{align}
For details, see Ref.~\cite{Sato:2019axv}.
In the limit of large $\tau$, $\varphi$ reaches a fixed point
and satisfies $\nabla^2 \varphi - \lambda[\varphi](\partial V/\partial\varphi)= 0$.
This fixed point is not a saddle point because we fixed ${\cal V}$,
and this property guarantees the stability of numerical calculation.
The bounce solution $\phi_B$ can be obtained by the following scale transformation of the fixed point $\varphi$.
\begin{align}
\phi_B(r) = \lim_{\tau\to\infty} \varphi(\lambda^{-1/2}r, \tau). \label{eq:bounce from gradient flow}
\end{align}
%
Note that ${\cal V}[\phi_B] < 0$ is guaranteed thanks to Eq.~(\ref{eq:initial V}) and Eq.~(\ref{eq:evolution V and T})
and $\phi_B$ is a configuration which is relevant with the false vacuum decay.

\section{Algorithm}
We solve the flow equation Eq.~(\ref{eq:gradient flow equation}) numerically.
We discretize $r$-space and take $n$ points at $r_i = (i-1) \delta r$ for $i=1,\cdots,n$.
The radius of the sphere is $R \equiv (n-1) \delta r$,
however, the value of $R$ itself is not important because we utilize the scale transformation property of the bounce solution.
We take the following discretized Laplacian which is similar to Ref.~\cite{Bar:2019bqz}:
\begin{align}
\nabla^2\varphi_i |_{r=r_j} = \left(\frac{d^2\varphi_i}{dr^2} + \frac{d-1}{r} \frac{d\varphi_i}{dr}\right)_{r=r_j}
=
\begin{cases}
\displaystyle\frac{2d(\varphi_{i,2} - \varphi_{i,1})}{\delta r^2} & (j=1) \\[4mm]
\displaystyle\frac{\varphi_{i,j+1} - 2\varphi_{i,j} + \varphi_{i,j-1}}{\delta r^2} + \frac{d-1}{r_j} \frac{\varphi_{i,j+1} - \varphi_{i,j-1}}{2\delta r} & (j>1) 
\end{cases}
\end{align}
The integrals in Eq.~(\ref{eq:lambda}) are approximated by the trapezoidal rule.
We take a boundary condition such that $\varphi_{i,n} = \phi_{FV,i}$.

Each step of the evolution of $\varphi$ is evaluated as
\begin{align}
\varphi_i(r,\tau+\delta\tau) = \delta\tau \times \left[
\nabla^2 \varphi_i - \lambda[\varphi] \frac{\partial V(\varphi)}{\partial \varphi_i}
\right]. \label{eq:evolution of phi}
\end{align}
We repeat this evolution until $\varphi$ converges.
For the stability of the numerical calculation, $\delta\tau$ should be ${\cal O}(\delta r^2)$
otherwise the fluctuation with large wave number exponentially grows.

We take the following configuration as the initial condition.
\begin{align}
\varphi_i(r,0) = \phi_{TV,i} + \frac{1}{2} (\phi_{FV,i} - \phi_{TV,i})\left( 1 + \tanh\left( \frac{ r_i - r_0 }{ \sigma } \right) \right).  \label{eq:initial condition}
\end{align}
Here $\phi_{TV}$ is chosen such that $V(\phi_{TV}) < V(\phi_{FV})$.
(Note that $\phi_{TV}$ is not necessarily to be the true vacuum.
Also the algorithm presented here works even there exists no true vacuum.
For example, Ref.~\cite{Sato:2019axv} shows this algorithm works with $V(\phi) = \phi^2/2-\phi^3/3$.)
The initial $r_0$ is set to some value which is smaller than $R$ ($0.5 R$ as default),
and $\sigma$ is set to be sufficiently small to obtain negative ${\cal V}$.
In order to satisfy the boundary condition at infinity, the first derivative of the field should be small enough.
If $\partial\phi_i / \partial r |_{r=R} = (\phi_{i,n} - \phi_{i,n-1})/\delta r$ is not small after solving the flow equation,
the size of the bounce is not small enough compared to the size of the space $R$.
In this case, we solve the flow equation again with smaller $r_0$.
See also Fig.~\ref{fig:flow chart}.

\tikzstyle{decision} = [diamond, draw,  
    text width=7.5em, text badly centered, node distance=3.5cm, inner sep=0pt]
\tikzstyle{block} = [rectangle, draw,
    text width=8em, text centered, rounded corners, minimum height=4em]
\tikzstyle{line} = [draw, -latex']
\begin{figure}
\centering
\begin{tikzpicture}[node distance = 2cm, auto]
	\node[block](init){Initialization and set $\tau$ as $0$};
	\node[block, below of=init](set initial){Take the configuration Eq.~(\ref{eq:initial condition})};
	\node[decision, below of=set initial](ask V){$V[\varphi]<0$?};
	\node[block, left of=ask V, node distance=5cm](change width){Take smaller $\sigma$};
	\node[block, below of=ask V, node distance=4cm](evolveUntil1){Evolve configuration by a small time step $\delta\tau$};
	\node[decision, below of=evolveUntil1](ask derivative){Are $\partial\varphi_i/\partial r$'s small enough?};
	\node[block, left of=ask derivative, node distance=5cm](change size){Take smaller $r_i$'s while fixing $\sigma/r_i$'s};
	\node[block, left of=evolveUntil1, node distance=5cm](set initial2){Take the configuration Eq.~(\ref{eq:initial condition}) and set $\tau$ as $0$};
	\node[decision, right of=ask derivative, node distance=4cm](evolveUntil2){$\tau<\tau_1$?};
	\node[block, below of=evolveUntil2, node distance=4cm](finish){Finish};
	\path [line] (init) -- (set initial);
	\path [line] (set initial) -- (ask V);
	\path [line] (change width) |- (set initial);
	\path [line] (ask V) -- node{Yes}(evolveUntil1);
	\path [line] (ask V) -- node{No}(change width);
	\path [line] (evolveUntil1) -- (ask derivative);
	\path [line] (ask derivative) -- node{No}(change size);
	\path [line] (change size) -- (set initial2);
	\path [line] (set initial2) -- (evolveUntil1);
	\path [line] (ask derivative) -- node{Yes}(evolveUntil2);
	\path [line] (evolveUntil2) -- node{No}(finish);
	\path [line] (evolveUntil2) |- node{~~~~~Yes}(evolveUntil1);
\end{tikzpicture}
\caption{Flow chart of the algorithm of \texttt{SimpleBounce}.
The default values of $\tau_0$ and $\tau_1$ are 0.05 and 0.4.
} \label{fig:flow chart}
\end{figure}
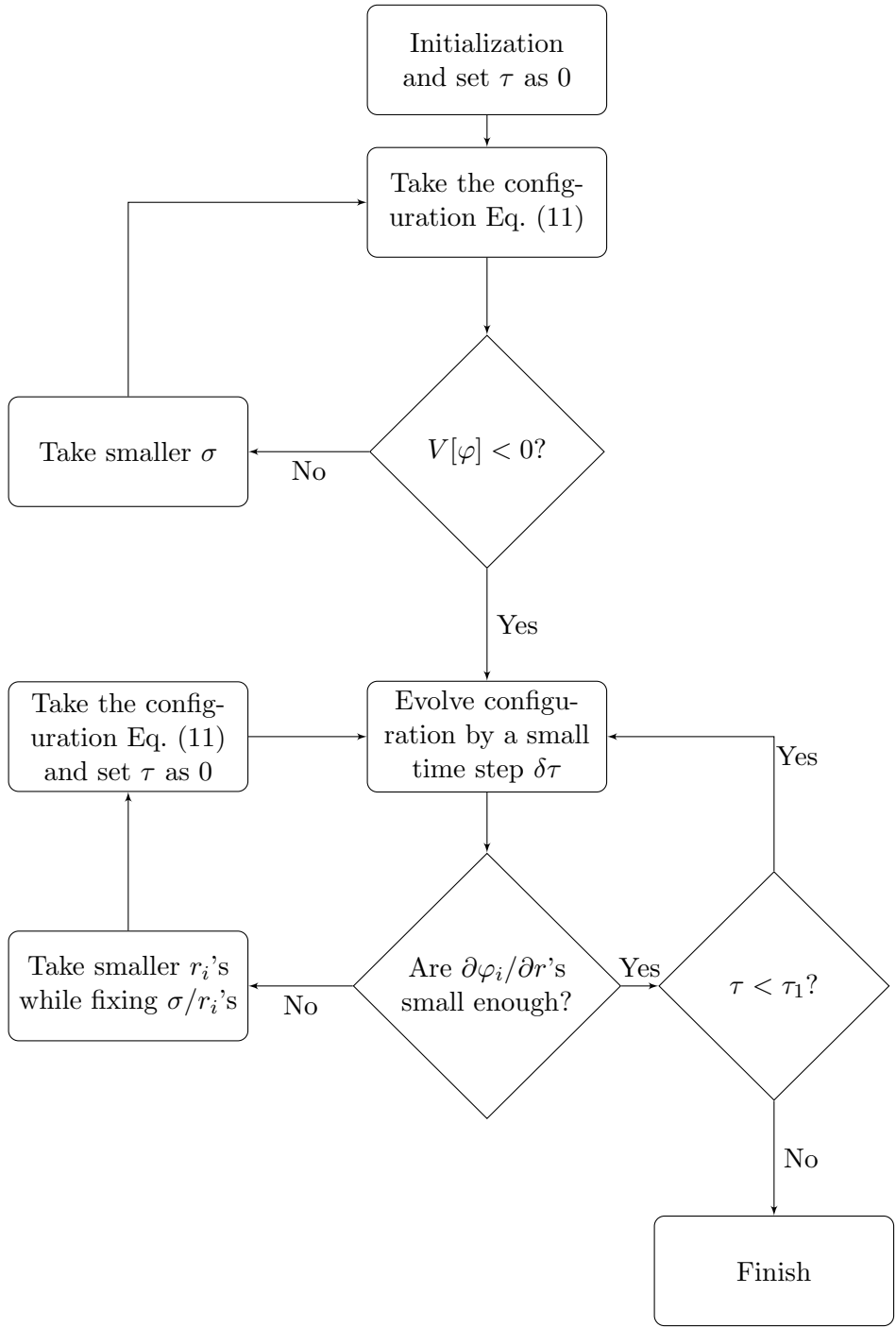

\section{User instructions}
This package is available at \url{https://github.com/rsato64/SimpleBounce}.
The main files are \texttt{simplebounce.cc} and \texttt{simplebounce.h}.
The sample executable files are compiled by
\begin{quote}
\texttt{make}
\end{quote}
The default compiler is \texttt{g++}. If you need to change this, please edit \texttt{Makefile.inc}.
In this package, the following examples are available.
\begin{itemize}
\item \texttt{sample1.x}\\
This calculates the Euclidean action for the bounce in a single scalar field model. See also section \ref{sec:singlescalar}.
\item \texttt{benchmark/compare\_with\_cosmotransitions/}\\
\texttt{run\_simplebounce.sh} in this folder gives Tab.~\ref{tab:performance}.
See also section \ref{sec:benchmarks}.
\item \texttt{benchmark/change\_n\_tau/}\\
\texttt{run.sh} in this folder gives Fig.~\ref{fig:accuracy}.
See also section \ref{sec:change n tau}.
\end{itemize}

\section{Sample codes and discussions}
Here we describe several sample codes and discuss their results.

\subsection{Sample code for single scalar field model}\label{sec:singlescalar}
Here we show an example of code.
The following information is required to calculate the bounce solution:
\begin{itemize}
\item The number of scalar field $n_\phi$
\item The scalar potential $V(\phi)$
\item The first derivative(s) of the scalar potential $\partial V/\partial\phi_i$
\item The position of the false vacuum $\phi_{FV}$
\item A point $\phi_{TV}$ which gives $V(\phi_{TV}) < V(\phi_{FV})$.
\end{itemize}

Here we take the potential as
\begin{align}
V(\phi) = \frac{1}{2}\phi^2 - \frac{1}{3}\phi^3.
\end{align}
The false vacuum is at $\phi=0$. The scalar field $\phi$ will tunnel into positive $\phi$ region.
The source file \texttt{sample1.cc} is given as
\begin{lstlisting}[basicstyle=\ttfamily\footnotesize, frame=single]
#include<iostream>
#include"simplebounce.h"
using namespace std;
using namespace simplebounce;

class MyModel : public GenericModel{
  public:
	MyModel(){
		setNphi(1); // number of scalar field(s)
	}
	// potential for scalar field(s)
	double vpot (const double* phi) const{
		return phi[0]*phi[0]/2. - phi[0]*phi[0]*phi[0]/3.;
	}
	// first derivative(s) of potential
	void calcDvdphi(const double *phi) const{
		dvdphi[0] = phi[0] - phi[0]*phi[0];
	}
};

int main() {

	BounceCalculator bounce;
	bounce.verboseOn(); // verbose mode
	bounce.setRmax(1.); // phi(rmax) = phi(False vacuum)
	bounce.setDimension(4); // number of space dimension
	bounce.setN(100); // number of grid
	MyModel model;
	bounce.setModel(&model);

	double phiTV[1] = {10.}; // a point at which V<0
	double phiFV[1] = {0.}; // false vacuum
	bounce.setVacuum(phiTV, phiFV);


	// calcualte the bounce solution
	bounce.solve();

	// show the results
	bounce.printBounce();

	// show the Euclidean action
	cout << "S_E = " << bounce.action() << endl;

	return 0;
}
\end{lstlisting}
Then, the executable file can be generated by, for example, \texttt{g++ sample1.cc simplebounce.cc -O3}.

\subsection{The Euclidean action for benchmark models}\label{sec:benchmarks}
Tab.~\ref{tab:performance} shows the Euclidean action and the runtime for several benchmark models,
which are taken from Refs.~\cite{Athron:2019nbd, Chigusa:2019wxb}.
We also show the results and runtimes by \texttt{CosmoTransitions} in the table.
We can see the results of our package agree with \texttt{CosmoTransitions}, and our package is faster for the benchmark models.
We can expect the running time is proportional to the number of the scalar field
because the code calculates Eq.~(\ref{eq:evolution of phi}) and Eq.~(\ref{eq:lambda}) at each time step.
Tab.~\ref{tab:performance} is roughly consistent with this expectation.

\begin{table}
\centering
\begin{tabular}{|c|cc|cc|}
\hline
& \multicolumn{2}{|c|}{${\cal S}_E$} & \multicolumn{2}{|c|}{Time[s]} \\
model & SB & CT & SB & CT \\\hline\hline
\#1 in Tab.~1 of \cite{Athron:2019nbd} & $52.4$             & $52.6$             & 0.04 & 0.05\\\hline
\#2 in Tab.~1 of \cite{Athron:2019nbd} & $20.8$             & $21.1$             & 0.04 & 0.35\\\hline
\#3 in Tab.~1 of \cite{Athron:2019nbd} & $22.0$             & $22.0$             & 0.07 & 0.17\\\hline
\#4 in Tab.~1 of \cite{Athron:2019nbd} & $55.8$             & $56.1$             & 0.07 & 0.31\\\hline
\#5 in Tab.~1 of \cite{Athron:2019nbd} & $16.2$             & $16.4$             & 0.09 & 0.26\\\hline
\#6 in Tab.~1 of \cite{Athron:2019nbd} & $24.4$             & $24.5$             & 0.11 & 0.25\\\hline
\#7 in Tab.~1 of \cite{Athron:2019nbd} & $36.6$             & $36.7$             & 0.14 & 0.22\\\hline
\#8 in Tab.~1 of \cite{Athron:2019nbd} & $45.9$             & $46.1$             & 0.16 & 0.23\\\hline
Eq.~40 of \cite{Chigusa:2019wxb}        & $1.08 \times 10^3$ & $1.09 \times 10^3$ & 0.03 & 0.09\\\hline
Eq.~41 of \cite{Chigusa:2019wxb}        & $6.62$             & $6.65$             & 0.03 & 0.06\\\hline
Eq.~42 of \cite{Chigusa:2019wxb}        & $1.75 \times 10^3$ & $1.77 \times 10^3$ & 0.34 & 0.54\\\hline
Eq.~43 of \cite{Chigusa:2019wxb}        & $4.45$             & $4.50$             & 0.05 & 0.19\\\hline
\end{tabular}
\caption{The comparison with \texttt{CosmoTransitions}.
For \texttt{CosmoTransitions}, we take \texttt{fRatioConv} as 0.02.
The bounce action ${\cal S}_E$ is calculated for $d=3$.
The runtimes are measured by Thinkpad X250 with Ubuntu 16.04,
whose CPU is Intel\textregistered Core\texttrademark i7-5600U (2.60 GHz) and compiler is GCC version 5.4.0.
We take \texttt{-O3} as the optimization option.
}\label{tab:performance}
\end{table}

\subsection{Changing $n$ and $\tau_1$}\label{sec:change n tau}
\begin{figure}
\centering
\includegraphics[width=0.8\hsize]{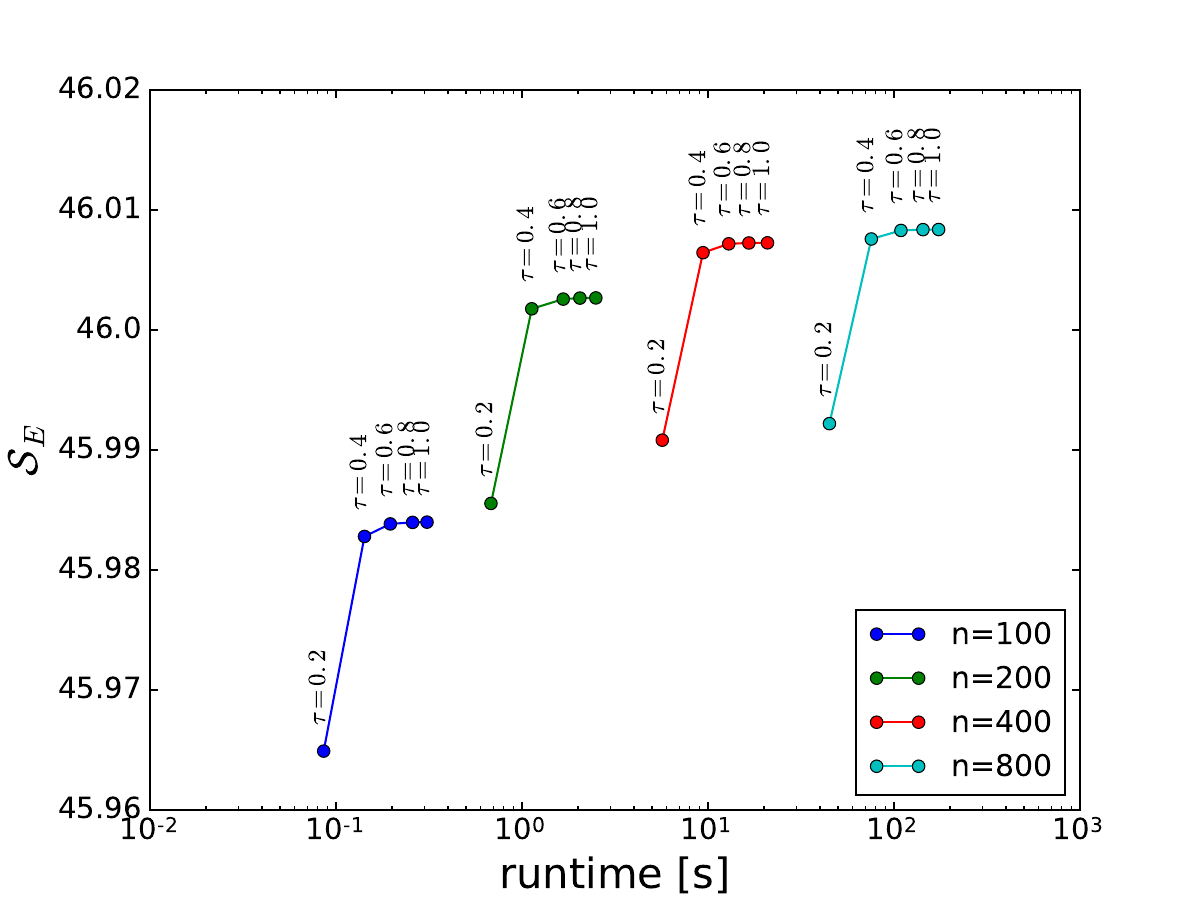}
\caption{The runtime of the value of ${\cal S}_E$ for different number of the lattice $n$ and the flow time $\tau_1$.
We take the model \#8 in Tab.~1 of \cite{Athron:2019nbd}.
The environment of PC is same as Tab.~\ref{tab:performance}.
}\label{fig:accuracy}
\end{figure}

In Fig.~\ref{fig:accuracy}, we show the value of ${\cal S}_E$ for different number of the lattice $n$ and the flow time $\tau_1$.
We can see that the value of ${\cal S}_E$ converges for large $n$ and large $\tau_1$,
and we can get the result of ${\cal O}(0.1)~\%$ accuracy in ${\cal O}(0.1)$~s run with $n=100$.
Fig.~\ref{fig:accuracy} also shows that the runtime is proportional to $n^3$.
This is because the step $\delta\tau$ is bounded as $\sim \delta r^2$.
Thus, in general, thin wall bounce takes much more time than thick wall bounce to obtain accurate result.

\section*{Acknowledgements}
The author thanks Tomohiro Abe, Yohei Ema, Parsa Ghorbani, Ryusuke Jinno, Thomas Konstandin, and Kyohei Mukaida for useful discussions,
and also thanks Yu Hamada for pointing out a typo in Eq.~(\ref{eq:bounce from gradient flow}).

\providecommand{\href}[2]{#2}\begingroup\raggedright\endgroup


\begin{thebibliography}{10}

\bibitem{Sato:2019axv}
R.~Sato, ``{Simple Gradient Flow Equation for the Bounce Solution},''
  \href{http://dx.doi.org/10.1103/PhysRevD.101.016012}{{\em Phys. Rev. D}
  {\bfseries 101} no.~1, (2020) 016012},
  \href{http://arxiv.org/abs/1907.02417}{{\ttfamily arXiv:1907.02417
  [hep-ph]}}.

\bibitem{Coleman:1977th}
S.~R. Coleman, V.~Glaser, and A.~Martin, ``{Action Minima Among Solutions to a
  Class of Euclidean Scalar Field Equations},''
\href{http://dx.doi.org/10.1007/BF01609421}{{\em Commun. Math. Phys.}
  {\bfseries 58} (1978) 211--221}.

\bibitem{Coleman:1977py}
S.~R. Coleman, ``{The Fate of the False Vacuum. 1. Semiclassical Theory},''
  \href{http://dx.doi.org/10.1103/PhysRevD.15.2929,
  10.1103/PhysRevD.16.1248}{{\em Phys. Rev.} {\bfseries D15} (1977)
  2929--2936}.
[Erratum: Phys. Rev.D16,1248(1977)].

\bibitem{Lee:1974ma}
T.~D. Lee and G.~C. Wick, ``{Vacuum Stability and Vacuum Excitation in a Spin 0
  Field Theory},''
\href{http://dx.doi.org/10.1103/PhysRevD.9.2291}{{\em Phys. Rev.} {\bfseries
  D9} (1974) 2291--2316}.

\bibitem{Frampton:1976kf}
P.~H. Frampton, ``{Vacuum Instability and Higgs Scalar Mass},''
  \href{http://dx.doi.org/10.1103/PhysRevLett.37.1378,
  10.1103/PhysRevLett.37.1716.2}{{\em Phys. Rev. Lett.} {\bfseries 37} (1976)
  1378}.
[Erratum: Phys. Rev. Lett.37,1716(1976)].

\bibitem{Profumo:2010kp}
S.~Profumo, L.~Ubaldi, and C.~Wainwright, ``{Singlet Scalar Dark Matter:
  monochromatic gamma rays and metastable vacua},''
  \href{http://dx.doi.org/10.1103/PhysRevD.82.123514}{{\em Phys. Rev.}
  {\bfseries D82} (2010) 123514},
\href{http://arxiv.org/abs/1009.5377}{{\ttfamily arXiv:1009.5377 [hep-ph]}}.

\bibitem{Wainwright:2011kj}
C.~L. Wainwright, ``{CosmoTransitions: Computing Cosmological Phase Transition
  Temperatures and Bubble Profiles with Multiple Fields},''
  \href{http://dx.doi.org/10.1016/j.cpc.2012.04.004}{{\em Comput. Phys.
  Commun.} {\bfseries 183} (2012) 2006--2013},
\href{http://arxiv.org/abs/1109.4189}{{\ttfamily arXiv:1109.4189 [hep-ph]}}.

\bibitem{Masoumi:2016wot}
A.~Masoumi, K.~D. Olum, and B.~Shlaer, ``{Efficient numerical solution to
  vacuum decay with many fields},''
  \href{http://dx.doi.org/10.1088/1475-7516/2017/01/051}{{\em JCAP} {\bfseries
  1701} no.~01, (2017) 051},
\href{http://arxiv.org/abs/1610.06594}{{\ttfamily arXiv:1610.06594 [gr-qc]}}.

\bibitem{Akula:2016gpl}
S.~Akula, C.~Bal\'{a}zs, and G.~A. White, ``{Semi-analytic techniques for
  calculating bubble wall profiles},''
  \href{http://dx.doi.org/10.1140/epjc/s10052-016-4519-5}{{\em Eur. Phys. J.}
  {\bfseries C76} no.~12, (2016) 681},
\href{http://arxiv.org/abs/1608.00008}{{\ttfamily arXiv:1608.00008 [hep-ph]}}.


\bibitem{Athron:2019nbd}
P.~Athron, C.~Balázs, M.~Bardsley, A.~Fowlie, D.~Harries, and G.~White,
  ``{BubbleProfiler: finding the field profile and action for cosmological
  phase transitions},'' \href{http://dx.doi.org/10.1016/j.cpc.2019.05.017}{{\em
  Comput. Phys. Commun.} {\bfseries 244} (2019) 448--468},
  \href{http://arxiv.org/abs/1901.03714}{{\ttfamily arXiv:1901.03714
  [hep-ph]}}.

\bibitem{Claudson:1983et}
M.~Claudson, L.~J. Hall, and I.~Hinchliffe, ``{Low-Energy Supergravity: False
  Vacua and Vacuous Predictions},''
\href{http://dx.doi.org/10.1016/0550-3213(83)90556-4}{{\em Nucl. Phys.}
  {\bfseries B228} (1983) 501--528}.

\bibitem{Kusenko:1995jv}
A.~Kusenko, ``{Improved action method for analyzing tunneling in quantum field
  theory},'' \href{http://dx.doi.org/10.1016/0370-2693(95)00994-V}{{\em Phys.
  Lett.} {\bfseries B358} (1995) 51--55},
\href{http://arxiv.org/abs/hep-ph/9504418}{{\ttfamily arXiv:hep-ph/9504418
  [hep-ph]}}.

\bibitem{Kusenko:1996jn}
A.~Kusenko, P.~Langacker, and G.~Segre, ``{Phase transitions and vacuum
  tunneling into charge and color breaking minima in the MSSM},''
  \href{http://dx.doi.org/10.1103/PhysRevD.54.5824}{{\em Phys. Rev.} {\bfseries
  D54} (1996) 5824--5834},
\href{http://arxiv.org/abs/hep-ph/9602414}{{\ttfamily arXiv:hep-ph/9602414
  [hep-ph]}}.

\bibitem{Moreno:1998bq}
J.~M. Moreno, M.~Quiros, and M.~Seco, ``{Bubbles in the supersymmetric standard
  model},'' \href{http://dx.doi.org/10.1016/S0550-3213(98)00283-1}{{\em Nucl.
  Phys.} {\bfseries B526} (1998) 489--500},
\href{http://arxiv.org/abs/hep-ph/9801272}{{\ttfamily arXiv:hep-ph/9801272
  [hep-ph]}}.

\bibitem{Cline:1998rc}
J.~M. Cline, J.~R. Espinosa, G.~D. Moore, and A.~Riotto, ``{String mediated
  electroweak baryogenesis: A Critical analysis},''
  \href{http://dx.doi.org/10.1103/PhysRevD.59.065014}{{\em Phys. Rev.}
  {\bfseries D59} (1999) 065014},
\href{http://arxiv.org/abs/hep-ph/9810261}{{\ttfamily arXiv:hep-ph/9810261
  [hep-ph]}}.

\bibitem{John:1998ip}
P.~John, ``{Bubble wall profiles with more than one scalar field: A Numerical
  approach},'' \href{http://dx.doi.org/10.1016/S0370-2693(99)00272-5}{{\em
  Phys. Lett.} {\bfseries B452} (1999) 221--226},
\href{http://arxiv.org/abs/hep-ph/9810499}{{\ttfamily arXiv:hep-ph/9810499
  [hep-ph]}}.

\bibitem{Cline:1999wi}
J.~M. Cline, G.~D. Moore, and G.~Servant, ``{Was the electroweak phase
  transition preceded by a color broken phase?},''
  \href{http://dx.doi.org/10.1103/PhysRevD.60.105035}{{\em Phys. Rev.}
  {\bfseries D60} (1999) 105035},
\href{http://arxiv.org/abs/hep-ph/9902220}{{\ttfamily arXiv:hep-ph/9902220
  [hep-ph]}}.

\bibitem{Konstandin:2006nd}
T.~Konstandin and S.~J. Huber, ``{Numerical approach to multi dimensional phase
  transitions},'' \href{http://dx.doi.org/10.1088/1475-7516/2006/06/021}{{\em
  JCAP} {\bfseries 0606} (2006) 021},
\href{http://arxiv.org/abs/hep-ph/0603081}{{\ttfamily arXiv:hep-ph/0603081
  [hep-ph]}}.

\bibitem{Park:2010rh}
J.-h. Park, ``{Constrained potential method for false vacuum decays},''
  \href{http://dx.doi.org/10.1088/1475-7516/2011/02/023}{{\em JCAP} {\bfseries
  1102} (2011) 023},
\href{http://arxiv.org/abs/1011.4936}{{\ttfamily arXiv:1011.4936 [hep-ph]}}.

\bibitem{Guada:2018jek}
V.~Guada, A.~Maiezza, and M.~Nemev\v{s}ek, ``{Multifield Polygonal Bounces},''
  \href{http://dx.doi.org/10.1103/PhysRevD.99.056020}{{\em Phys. Rev.}
  {\bfseries D99} no.~5, (2019) 056020},
\href{http://arxiv.org/abs/1803.02227}{{\ttfamily arXiv:1803.02227 [hep-th]}}.

\bibitem{Jinno:2018dek}
R.~Jinno, ``{Machine learning for bounce calculation},''
\href{http://arxiv.org/abs/1805.12153}{{\ttfamily arXiv:1805.12153 [hep-th]}}.

\bibitem{Espinosa:2018szu}
J.~R. Espinosa and T.~Konstandin, ``{A Fresh Look at the Calculation of
  Tunneling Actions in Multi-Field Potentials},''
  \href{http://dx.doi.org/10.1088/1475-7516/2019/01/051}{{\em JCAP} {\bfseries
  1901} no.~01, (2019) 051},
\href{http://arxiv.org/abs/1811.09185}{{\ttfamily arXiv:1811.09185 [hep-th]}}.

\bibitem{Piscopo:2019txs}
M.~L. Piscopo, M.~Spannowsky, and P.~Waite, ``{Solving differential equations
  with neural networks: Applications to the calculation of cosmological phase
  transitions},'' \href{http://dx.doi.org/10.1103/PhysRevD.100.016002}{{\em
  Phys. Rev.} {\bfseries D100} no.~1, (2019) 016002},
\href{http://arxiv.org/abs/1902.05563}{{\ttfamily arXiv:1902.05563 [hep-ph]}}.

\bibitem{Chigusa:2019wxb}
S.~Chigusa, T.~Moroi, and Y.~Shoji, ``{Bounce Configuration from Gradient
  Flow},'' \href{http://dx.doi.org/10.1016/j.physletb.2019.135115}{{\em Phys.
  Lett. B} {\bfseries 800} (2020) 135115},
  \href{http://arxiv.org/abs/1906.10829}{{\ttfamily arXiv:1906.10829
  [hep-ph]}}.

\bibitem{lopes1996radial}
O.~Lopes, ``Radial symmetry of minimizers for some translation and rotation
  invariant functionals,'' \href{http://dx.doi.org/10.1006/jdeq.1996.0015}{{\em
  Journal of differential equations} {\bfseries 124} no.~2, (1996) 378--388}.

\bibitem{byeon2009symmetry}
J.~Byeon, L.~Jeanjean, and M.~Mari{\c{s}}, ``Symmetry and monotonicity of least
  energy solutions,'' \href{http://dx.doi.org/10.1007/s00526-009-0238-1}{{\em
  Calculus of Variations and Partial Differential Equations} {\bfseries 36}
  no.~4, (2009) 481--492}, \href{http://arxiv.org/abs/0806.0299}{{\ttfamily
  arXiv:0806.0299 [math.AP]}}.

\bibitem{Blum:2016ipp}
K.~Blum, M.~Honda, R.~Sato, M.~Takimoto, and K.~Tobioka, ``{O($N$) Invariance
  of the Multi-Field Bounce},''
  \href{http://dx.doi.org/10.1007/JHEP05(2017)109,
  10.1007/JHEP06(2017)060}{{\em JHEP} {\bfseries 05} (2017) 109},
  \href{http://arxiv.org/abs/1611.04570}{{\ttfamily arXiv:1611.04570
  [hep-th]}}.
[Erratum: JHEP06,060(2017)].

\bibitem{Bar:2019bqz}
N.~Bar, K.~Blum, J.~Eby, and R.~Sato, ``{Ultralight dark matter in disk
  galaxies},'' \href{http://dx.doi.org/10.1103/PhysRevD.99.103020}{{\em Phys.
  Rev.} {\bfseries D99} no.~10, (2019) 103020},
\href{http://arxiv.org/abs/1903.03402}{{\ttfamily arXiv:1903.03402
  [astro-ph.CO]}}.

\end{thebibliography}
\end{document}